\documentclass[useAMS,usenatbib]{mn2e}
\usepackage{color}
\def\apj{{\it Astrophys.~J.}}  
\def\apjl{{\it Astrophys.~J.~Lett.}}  
\def\apjs{{\it Astrophys.~J.~Suppl.}}  
\def\mnras{{\it Mon. Not. R. astr. Soc.}}      
\def\aap{{\it Astron.~Astrophys.}}     
\def\araa{{\it ARA\&A}}    
\def\physrep{{\it Phys.~Rep.}} 
\def\nat{{\it Nature}} 
\def\aj{{\it Astron.~J.}} 
\title[SMGs: two populations and redshift cut-off]{The evolution of
submillimetre galaxies: two populations and a redshift cut-off}
\author[J. V. Wall, A. Pope, Douglas Scott]{J. V. Wall\thanks{E-mail:
jvw@astro.ubc.ca}, Alexandra Pope and Douglas Scott\\
Physics and Astronomy Department, University of British Columbia, 6224
Agricultural Road, Vancouver, V6T 1Z1, Canada}
\begin{document}

\setlength{\topmargin}{2mm}

\date{2007 September 22}


\maketitle


\begin{abstract}
We explore the epoch dependence of number density and star-formation rate for
submillimetre galaxies (SMGs) found at $850\,\mu$m. The study uses a sample of 38 SMG in
the GOODS-N field, for which cross-waveband identifications have been obtained for
35/38 members together with redshift measurements or estimates. A
maximum-likelihood analysis is employed, along with the `single-source-survey'
technique. We find a diminution in both space density and star formation rate at
$z>3$, closely mimicking the redshift cut-offs found for QSOs selected in
different wavebands. The diminution in redshift is particularly marked, at a
significance level too small to measure. The data further suggest, at a
significance level of about 0.001, that two separately-evolving populations may be
present, with distinct luminosity functions. These results parallel the different
evolutionary behaviours of LIRGs and ULIRGs, and represent another manifestation
of `cosmic down-sizing', suggesting that differential evolution extends to the
most extreme star-forming galaxies.

\end{abstract}


\section{Introduction}

`Submillimetre galaxies' (SMGs) represent a major population of massive
star-forming galaxies at high redshift (e.g. \citealt{hug98, bla02}). Found in
limited-area sky surveys at $850\,\mu$m with the Submm Common-User Bolometric
Array (SCUBA; \citealt{hol99}) on the James Clerk Maxwell Telescope, they are
believed to be dust-enshrouded starbursts, with the dust heated by UV radiation
from young stars. They may be the distant early equivalents of the local
prodigious star-formers, the ULIRGs and LIRGs (`Ultra-Luminous IR Galaxies',
`Luminous IR Galaxies', \citealt{san96}). SMGs appear to carry much of the
star-formation rate density (SFRD) of the early Universe on their shoulders.
Understanding these objects is thus fundamental to our understanding of galaxy
formation. Several attempts have been made to track their contribution to the
global SFRD as a function of epoch \citep[e.g.][]{lil99,cha05}. These efforts have
been hampered by incomplete cross-waveband identifications and hence the
subsamples which have redshifts have been biased.  Recently, \citet{pop05,pop06}
succeeded in identifying 35 out of a sample of 38 SMGs from the SCUBA survey in
the GOODS(Great Observatories Origins Deep Survey)-N field, and secured redshift
estimates for all identifications. It is the object of this paper to use this
sample to form a picture of the space density of these SMGs, and of the epoch
dependence of both their space density and the corresponding SFRD.

The distinctive feature of the spectral energy
distributions of SMGs is the dominance of the cold-dust
spectrum, approximately that of a $\beta\simeq1.5$
greybody at $35\,$K,  peaking (rest frame $\nu I_\nu$) at
frequencies near $3200\,$GHz, wavelengths near $90\,\mu$m.
Such a spectrum implies that the K-correction is generally
negative (i.e.~we ascend the Rayleigh-Jeans tail as the
object moves to higher redshifts), and that there is
nothing to stop such objects being visible out to
redshifts of 5 or more \citep[e.g.][]{bla93}.
However, the redshift distribution of SMGs appears to peak around $\sim$ 2.2,
with little high-redshift tail beyond 4
\citep{cha05,pop06}, so that there is qualitative evidence
for a redshift diminution at early epochs. One of the aims
of this paper is to quantify this diminution.

The wide-spread phenomenon of `cosmic down-sizing' appears to be at variance with
the idea of hierarchical build-up of galaxies.  The emerging picture is that
although dark matter haloes build up hierarchically, the behaviour of the baryons
within these haloes is much more complicated.  In cosmic down-sizing, the dominant
activity becomes carried by more numerous, lower-luminosity, lower-mass objects at
progressively later times. The `down-sizing' \citep{cow96} originally described
how dominant star formation in galaxies shifted from luminous rare galaxies at
earlier epochs to more numerous and less luminous galaxies at recent epochs. In
addition to star formation, cosmic down-sizing is now known to apply to other
phenomena: AGN activity in X-ray QSOs \citep{ued03} and in radio galaxies (where
it has been known for 40 years in the guise of `differential evolution';
\citealt{lon66}), and SFR in ULIRGs+LIRGs \citep{per05,lef05,cha06}. This paper
examines whether the concept further extends to SMGs and the `cold dust' star
formation rate (SFR) associated with them.

\section{The submillimetre sample}

\subsection{The GOODS-N supermap}

Currently, the largest SMG sample which is almost
completely identified is from the GOODS-N field: all SCUBA
data from several extensive imaging campaigns in the
GOODS-N field have been combined into one 850-$\mu$m map,
referred to as the `supermap' (see \citealt{bor03} and
references therein). This supermap has noise properties
that vary strongly with position, but this can be
accounted for in the source extraction procedure.

The most recent published version of the supermap contains 35 $850\,\mu$m sources
detected above $3.5\sigma$ and satisfying a flux `de-boosting' threshold
\citep{pop06}. Subsequent to this work, the inclusion of additional two-bolometer
chopping photometry data in the supermap has resulted in three new $850\,\mu$m
sources, two of which have secure identifications. The sample therefore totals 38;
description of changes to the supermap as a result of these new data together with
identification of the three sources is in \citet{pop07}. Appendix~A gives the
relevant data for these three sources.

It is important in the present context to be clear as to why the GOODS-N supermap
yields a complete and unbiased sample. The issue is
covered in previous papers (\citealt{pop06} and references therein), and we
reiterate and summarize the points here.

The supermap is in fact a maximum-likelihood estimate of the flux of a point
source centered on each pixel. We take all the available data, from whatever
observing mode, to obtain this estimate.  To do this we use the beam-shape, as
well as the chop information, as discussed in previous papers \citep{bor03,pop05}.
The beauty of this approach is that we can include jiggle-maps with different
chops as well as the scan-map, and we can also fold in photometry data in exactly
the same way. Several checks were carried out to show that the different estimates
for specific sources agree with each other, and that the statistics of the data
going into each pixel are well behaved. There is no evidence of any bias from this
procedure, except for the usual flux boosting, which occurs when a signal-to-noise
threshold is applied to low S/N source data drawn from a steep source count. To
account for this, we followed the Bayesian approach of \citet{cop05}, which
reduces to a correction to the signal and noise of a source, dependent separately
on that signal and noise.  This method also has been extensively tested.  The
procedure allows an estimate of the fraction of sources which may not be real
(subject to some interpretation of what constitutes a `source' when approaching
the confusion limit), and the expectation is that there are only 1 or 2 such
objects. The de-boosting procedure led to removal of a handful of objects, most of
which are in fact likely to be real, but for each of them the chances of being
simply a noise excursion is not small. These are precisely the sorts of
`$3-3.5\sigma$' sources with relatively high noise which plagued earlier SMG
follow-up work. After such sources are removed the final catalogue is very
reliable [as we have confirmed with AzTEC (Astronomical Thermal Emission Camera;
Perera et al., in preparation) and MAMBO (Max Planck Millimetre Bolometer; Greve
et al., in preparation) maps of the GOODS-N field].

The resulting 38 sources, we believe constitute the most carefully compiled and
complete sample of submm objects currently available.

With regard to comparing the sources in our list with the results obtained for
radio-detected SMGs in GOODS-N by \citet{cha05},  our view is that these are
complementary studies.  The Chapman et al. sample relies heavily on pointed
photometry towards optically-faint radio sources only. The resulting data have
been extremely useful for understanding the properties of some SMGs, but
unfortunately the sample is hard to use in a statistical sense, because there is
no way to assess how biased it is, how complete it is, or what fraction of the
sources might be interlopers.

As an example of how all data were included in the supermap, some of our sources,
including two of the of the new ones, are from pointed photometry data. However,
what we have done is to include all of these data in the new super-map, treating
these data in precisely the same way as all the others, using the same source
extraction threshold and de-boosting procedure.  The main point is that this
process includes all of the bolometers, and not just the central one, so that what
we are effectively doing is adding under-sampled images to parts of the super-map.
Given that there are 37 bolometers, the fact that the central one happened to be
pointed at a specified place is not particularly relevant. These additional
photometry data simply help the S/N in the supermap in an inhomogeneous way, which
is easily tracked through estimation of the accompanying noise map.  That we do
not recover some of the \citet{cha05} sources is just because they fail our source
extraction criteria -- for an individual pointed observation, it may be reasonable
to take $3\sigma$ or less as a detection threshold, but such low values cannot be
used when examining an entire map.

\subsection{Identifications: the final sample}

Using the multi-wavelength data available in GOODS-N, likely counterparts were
found for $35/38$ of the SMGs. Full details on the identification process can be
found in \citet{pop06}. In brief: we searched for counterparts to the SMGs within
a search radius of 8 arcseconds using primarily the radio, Spitzer MIPS
(Multi-band Imaging Photometer) and IRAC (Infrared Array Camera) data. A
counterpart is considered secure if it has a probability of random association,
$P$, less than 5 per cent. The probabilities of random association are given in
Table A2 of \citet{pop06} with the exception of the two new identifications. These
two new identifications (Table~\ref{tab_newdata}), GN39 and GN40, have
random-association probabilities of 0.02 and 0.003, respectively, and therefore
qualify as secure. In fact 23/35 sources have secure ($P<0.05$) counterparts and
we thus expect that at most one of these will be an incorrect identification. The
rest of the counterparts, 12/35, are less secure, with $0.05<P<0.2$. However, when
we combine the probabilities of these less secure identifications, we expect only
1.3 incorrect identifications amongst them. For our total sample of 35 SMGs
identifications, we therefore expect at most 2--3 incorrect counterparts. As we
describe in \S\ref{explore} and \ref{two}, we have folded this uncertainty into
our analysis.

The extensive optical and infrared data yielded reliable photometric redshifts in
the absence of spectroscopic redshifts \citep{pop06}; 17 SMGs in the sample have
spectroscopic redshifts and the remaining 18 have optical or {\it Spitzer}
photometric redshift estimates. Optical photometric redshifts of these SMGs were
shown to be accurate by Pope et al.~(2005). Pope et al.~(2006) used the {\it
Spitzer} photometry to derive  model-independent estimates of the redshifts for
sources without reliable optical photometric redshifts (9/35 sources). These
redshifts were found to be accurate to $\sigma(\Delta z/(1+z)) =0.07$.

This sample of 35 objects is the basis for the following analysis of space density.
Throughout, we use a concordance cosmology with $\Omega_{\rm tot}=1.0$,
$\Omega_{\rm m}=0.3$, $\Omega_\Lambda=0.7$ and $h = 0.7$.

\section{Exploring space distribution}
\label{explore} For each SMG, we calculated the specific submm luminosity (at rest
frame, $850\,\mu$m) using only the $850\,\mu$m flux and the redshift, and assuming
a greybody spectral energy distribution with emissivity $\beta=1.5$ and dust
temperature $T=35\,$K. While there will be some scatter in $T$ and $\beta$, these
values provide a good description of the data as found in a number of submm
surveys \citep{cha05,kov06,pop06}.

\begin{figure}
\vspace{5.2cm}

\includegraphics{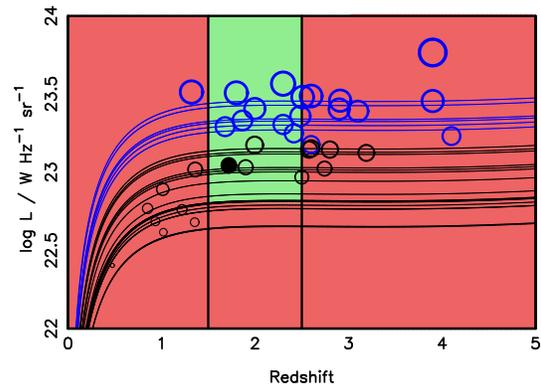}

\caption{The $L\,{-}\,z$ plane for all 35 SMGs. The curved lines represent the 35
different survey cut-offs for these objects; every one of the objects lies above
its cut-off line and these cut-off lines differ because of the differences in
local noise properties in the supermap.  Dividing the sample at the median value
in log($L$) (see \S\ref{two}), black lines /dots represent the lower-luminosity
objects; blue lines/dots represent the higher-luminosities, and dot size is
representative of log($L$). Note the remarkable form of these cut-off lines; some
of these objects can be seen out to effectively infinite redshifts because of the
inverse K-correction. This plot already suggests the basic result -- a dearth of
luminous sources below $z \simeq 1.5$ {\it and\/} a dearth of weaker sources above
$z \simeq 3$. The plot shows also how the `single-source-survey' technique works.
Taking the solid dot as a representative source, the heavy curve represents its
individual survey cut-off, and in this instance the calculation of space density
is for a redshift range 1.5 to 2.5. The function $\Omega_i(l,z)$ is shown as the
green area, over which it has a constant value equal to the individual area
relevant to the single source (see text); the function is zero (red) elsewhere.}

\label{Pz-submm1}

\end{figure}

The objects are shown in a luminosity--redshift
($L\,{-}\,z$) plot in Fig.~\ref{Pz-submm1}. It is clear
from this figure that some standard luminosity function
analyses will not work. The unique geometry means that the
$1/V_{\rm max}$ method in particular is problematic,
because most sources `see' no survey limit. Moreover,
beyond establishing the reality of evolution or otherwise,
the $1/V_{\rm max}$ method is poorly suited to small
samples. We therefore adopt a maximum-likelihood approach,
as first advocated by \citet{mar83} and used recently
in the detailed analysis of X-ray QSOs by \citet{ued03}.

Thus, following \citet{mar83}, consider the sample as a single homogeneous set of
$i$ objects, for which $\rho(z,L)(\partial V/\partial z) dz dL$ is the number in
volume element $(\partial V/\partial z)dz$ in luminosity element $dL$. The sky
fraction $\Omega_i(z,L)$ accessible to each object $i$ is unique -- each of our
objects is: (a) observable over an area of different physical size; and (b) has
its own flux-density limit line in the $L\,{-}\,z$ diagram. This factor
$\Omega_i(z,L)$ is thus essential in introducing the feature of the
single-source-survey \citep{wal05} by which each object is treated as having
unique access to the $L\,{-}\,z$ plane (Fig.~\ref{Pz-submm1}). The treatment is
analogous to the final survey having been done as 35 individual surveys finding a
single source each. The unique area accessible to each object on the $L\,{-}\,z$
plane is multiplied by its unique effective survey area to determine the final
value of its $\Omega_i(z,L)$. This effective survey area is a function solely of
flux density, with the relation as determined by \citet{bla06}.

The $\cal L$(ikelihood) function for the $i^{\rm th}$ object
is the probability of observing {\it one} object in its
$(dz,dl)$ element times the probability of observing {\it
zero} objects in all other $(dz,dl)$ elements accessible
to it. The Poisson model is the obvious one for the
likelihood:
\begin{equation}f(x:\mu)=\frac{{\rm e}^{-\mu}\mu^x}{x!},\end{equation}
where $\mu$ is the expected number. If $x=1$, the function is $\mu {\rm e}^{-\mu}$
and if $x=0$ it is ${\rm e}^{-\mu}$.

With $\rho(z,L)$ as the full description of space density,
\begin{equation} \mu=\lambda(z,L)\,dz\,dL,\ \ \mbox{for}\ \
\lambda=\rho(z,L)\Omega(z,L)(\partial V/\partial z).
\end{equation}

\noindent Hence
\begin{equation}
{\cal L} = \prod_{i}^{N} \lambda(z_i,L_i)\,dz dL\,e^{-\lambda(z_i,L_i)\,dz\,dL}
\prod_{j \neq i}^{N} e^{-\lambda(z_j,L_j)\,dz\,dL}\!,
\end{equation}

\noindent where $i$ denotes the elements of the $(z,L)$ plane in
which SMGs are present and $j$ denotes all others. From this, if
$S=-2\,\ln \cal{L}$, then
\begin{eqnarray}
S & = &-2\sum_{i=1}^N\,\ln \rho(z_i,L_i) \nonumber \\
&&+ \sum_{i=1}^N \int_z\int_L \rho(z,L) \Omega_i(z,L)
\frac{\partial V}{\partial z}\,dz\,dL + \mbox{constant.}
\end{eqnarray}

Consider simple factorizable density evolution of the form
$\rho(L,z) = \rho(z{=}0,L) \cdot \phi(z)$. In this
formulation we adopt a power-law luminosity function,
\begin{equation}
    \frac{dN}{dL} = \rho(L,z) = \frac{\rho_0}{L_*} \, \phi(z) \,
    \left( \frac{L}{L_*} \right) ^{-\alpha}\!.
\end{equation}
With $l\equiv L/L_*\,$, we have the local luminosity
function as $\rho(z{=}0,L) = (\rho_0/L_*)\,l^{-\alpha}$.
For the evolution function we again adopt a power-law,
$\phi(z)=(1+z)^k$.

If we substitute these assumptions into equation~(4) and set the derivative with
respect to $\rho_0$ to zero, we get a maximum-likelihood estimate for $\rho_0$:
\begin{equation}
    \rho_0 =
    \frac{N}{\sum_{i=0}^N \int_z\int_l
     (1+z)^k\,l^{-\alpha}\,\Omega_i(z,l)\,(\partial V /
    \partial z) \, dz \, L_*dl}.
\end{equation}
Putting this back into equation~(4) gives
\begin{eqnarray}
S & = & -2\sum_i^N \ln[(1+z_i)^k\,l_i^{-\alpha}] \nonumber \\
&& +2N\,\ln\sum_i^N  \int_z\int_l
(1+z)^k\,l^{-\alpha}\,\Omega_i(z,l)\,\left(\frac{\partial
V}{\partial z}\right)\,dz\,dl \nonumber \\
&& + (2N-2N\,\ln N).
\end{eqnarray}





\begin{figure}
\vspace{10.8cm}

\includegraphics{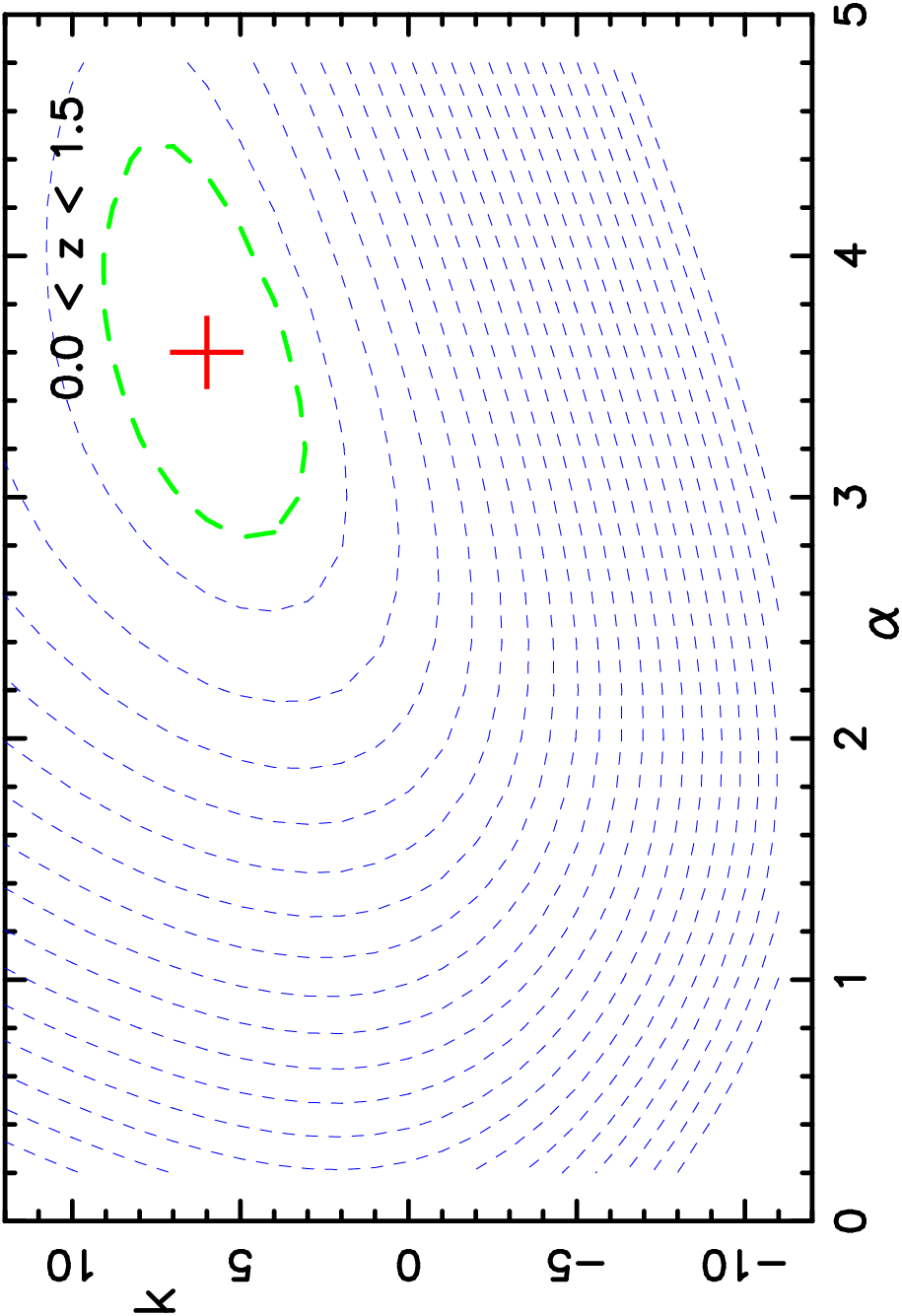}

\includegraphics{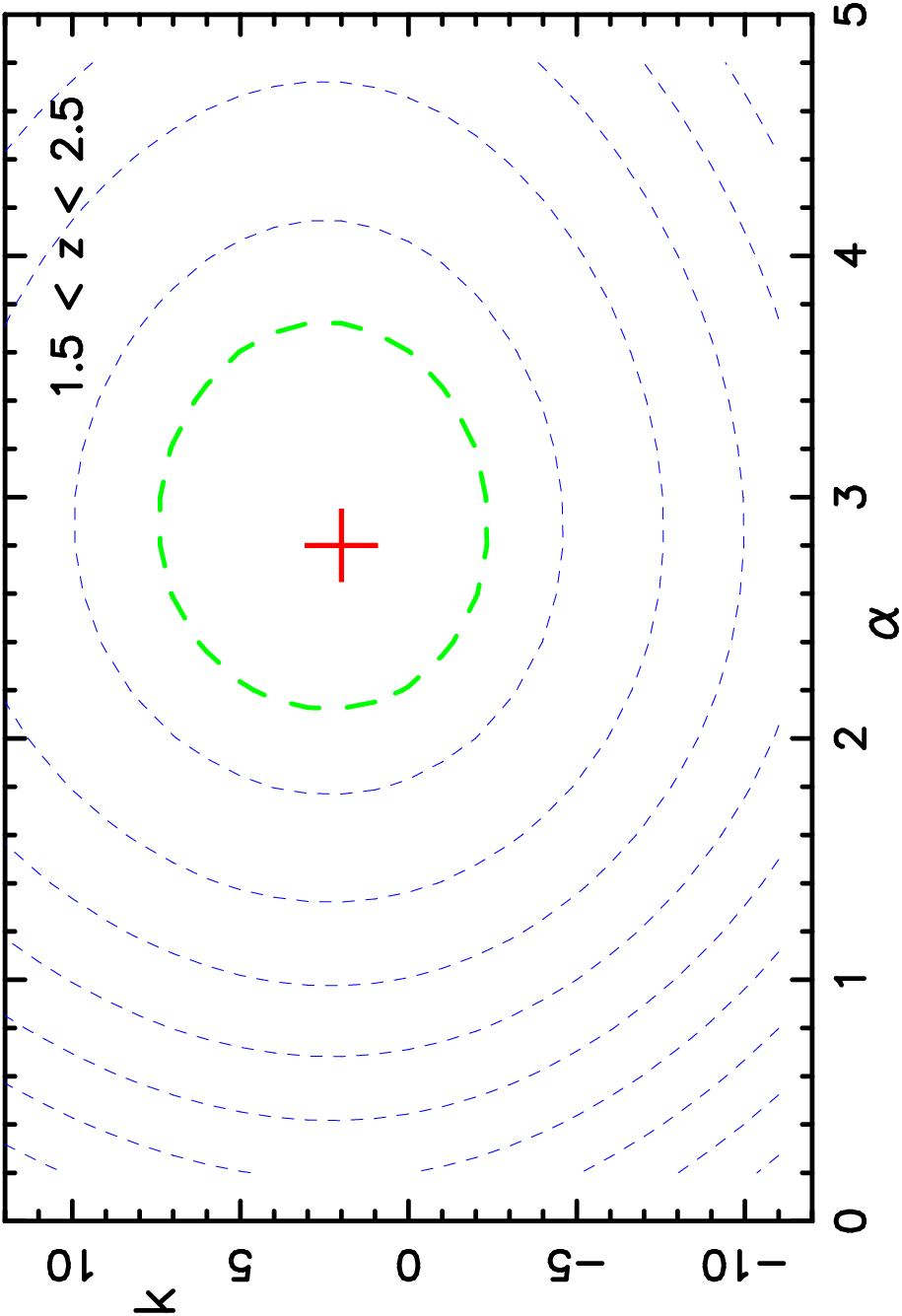}

\includegraphics{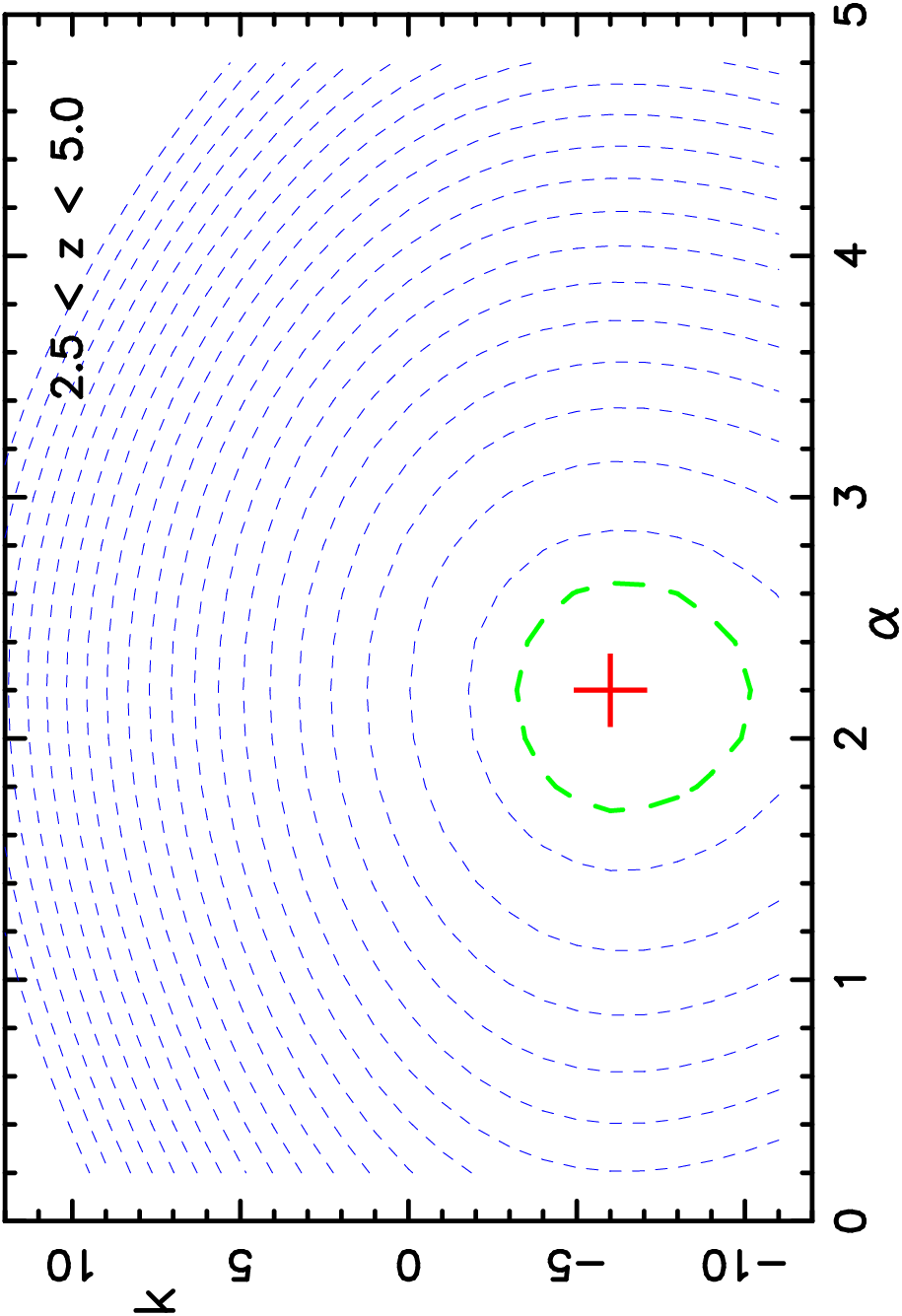}



\caption{Contours of the likelihood function $S$ for the parameters $k$ vs
$\alpha$ (evolution exponent vs luminosity-function slope) for broad redshift
slices. Best-fit values are indicated by red crosses, while green contours show
the $1\sigma$ uncertainty. The best-fit values of $k$ in the three plots indicate
rapid positive evolution in space density for $0<z<1.5$, a plateau at $1.5<z<2.5$,
and severe negative evolution or diminution at $2.5<z<5.0$.}

\label{z-slices}

\end{figure}

Inspection of Fig.~\ref{Pz-submm1} shows immediately that a single-power-law
function to describe density evolution will not work.  The density of points
clearly rises with increasing redshift before $z=2$ and falls after $z=3$.
Accordingly, we calculated the value of this likelihood function using a grid in
($k,\alpha$) for broad slices in redshift, with results shown in
Fig.~\ref{z-slices}. This figure shows that: (a) the slope of the luminosity
function does not change drastically with redshift; and (b) $k$, the $(1+z)$
exponent, changes from values around 5 at redshifts $<1.5$, to about zero for
$1.5<z<2.5$, to negative values at $z>2.5$.

\begin{figure}
\vspace{5.0cm}

\includegraphics{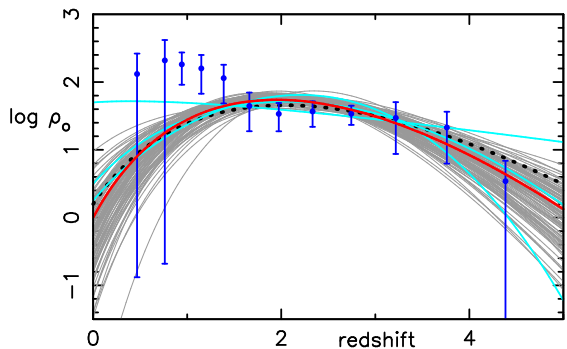}

\caption{Relative space density in successive redshift slices of width $\Delta\log
z = 0.7$, incrementing each bin mid-point by $0.25$ in $\log z$ (see the text).
Error bars are derived from $\sqrt{N}$, with $N$ the number of objects per bin.
The red curve is a maximum-likelihood functional fit to all the data.  Grey lines
represent 100 bootstrap trials of the evolution model, with bootstrapping
end-to-end from the initial sample of 35 SMGs. The three light blue curves
represent exponential cut-off models (described in the text); the uppermost
corresponds to $n=1$. The dotted line is the result of including the three
unidentified sources and ascribing them each a redshift of 4.0 (see the text).
Note that the curves are not fits to the points in the diagram -- they result from
the best likelihood fit to the entire luminosity--redshift plane for the assumed
luminosity-function form.}

\label{evol}

\end{figure}

We then used this simple formulation of the evolution function as follows --
ascribe {\it zero\/} evolution ($k=0.0$) across individual narrow redshifts
slices, adopt a (best) single-valued power law for the luminosity function
($\alpha=2.5$), and calculate the maximum-likelihood value of $\rho_0$, the
normalization of this luminosity function in each slice. The results should
roughly map space density with epoch, and are shown in Fig.~\ref{evol}.

This evolution of the luminosity function with
redshift was examined with a simple modification of the previous
density evolution: replacing the original power of $(1+z)$, namely
$k$, with the modified power $(k+\gamma z)$, i.e.
\begin{equation}
\rho(L,z)=\rho_o(1+z)^{(k+\gamma z)}l^{-\alpha}.
\end{equation}
There is as little physical justification for introduction of the $\gamma z$ term
in the exponent as there was for the assumption of the initial power law, or for
the factorization. However, the term provides a generic description of redshift
behaviour -- if $\gamma$ is negative, there is a roll-off in density toward higher
$z$. The results are again shown in Fig.~\ref{evol}. The red curve is a
minimization of the likelihood function $S$ for all three parameters $k$, $\gamma$
and $\alpha$, determined with a downhill simplex routine \citep{pre92}. The
maximum likelihood was found at $k=6.0 \pm 2.5$, $\gamma = -1.2 \pm 0.4$, and
$\alpha=2.5 \pm 0.3$ (see Table~\ref{tablum}).  The curves describe the individual
slice normalizations reasonably. The exponent of the initial rise ($k$) is similar
to those found in investigations of objects at other frequencies (radio and X-ray
AGN; see e.g. \citealt{wal80}); the roll-off shows a maximum space density of the
SMGs at about $z=2.0$, in accordance with the appearance of the $L{-}z$ plane
(Fig.~\ref{Pz-submm1}). Fig.~\ref{evol} includes data from a bootstrap analysis;
100 end-to-end bootstrap results from the original sample of 35 objects are shown.
Some 200 were done in all and none produced a value of $\gamma$ approaching zero.

\begin{table*}
\begin{center}
\caption{Best-fit density evolution parameters.}

\begin{tabular}{crlll}

\hline\hline
(Sub) sample$^{\ddagger}$ & $\rho_0$/Gpc$^{-3}$&\ \ \ $\alpha$ &\ \ \ $k$ &\ \ \ \ $\gamma$ \\

\hline
              &       &            &           &             \\

complete (35) & 3490 &  2.5 (0.3) & 6.0 (2.5) & $-1.2$ (0.4)\\
              &       &            &           &             \\
complete+3* (38) & 6334 & 2.5     & 4.8       & $-0.9$      \\
              &       &            &           &             \\
T$_{\rm eff}=35\pm10$K (35)& 7685 & 2.4 & 4.4  & $-0.9$      \\
              &       &            &           &             \\
T$_{\rm eff}=10(1+z)$K (35)&  2956 & 2.4 & 5.6 & $-1.1$      \\
              &       &            &           &             \\
$z'=z\pm0.14(1+z)$ (35) & 14010     & 2.5 & 3.7 & $-0.8$      \\
              &       &            &           &             \\
low-$L^\dag$ (17) & 11970 &  2.1 (0.8) & 5.3 (3.0) & $-1.3$ (0.7)\\

              &           &          &     &          \\

high-$L^\dag$ (18) &  208 & 3.3 (0.7) & 12.3 (6.0) & $-1.8$ (0.7)\\

              &           &          &     &          \\

\hline

\end{tabular}
\label{tablum}
\end{center}

$^{\ddagger}$Sample/sub-sample number of sources in brackets.\ \ \ \ \ \ \ \ \ \ \ \ \ \ \ \ \ \ \ \ \ \ \ \ \ \ \ \ \ \ \ \ \ \ \ \ \ \ \ \ \ \ \ \ \ \ \ \ \ \\
$^*$Total sample, with redshifts of 4.0 assigned to the three
unidentified sources.\ \ \ \ \ \ \ \ \ \ \ \ \ \ \ \\
$^\dag$Two sub-populations, with the sample of 35 divided at the
median $850\,\mu$m luminosity,\\
 $\log(L_{850\,\mu{\rm m}}/{\rm W}\,{\rm Hz}^{-1}{\rm sr}^{-1})
= 23.2$.
\end{table*}

A tenet of Bayesian analysis is that all obvious models should be tried. We
considered pure luminosity evolution, but on the assumption of a power-law
luminosity function, this is identical to density evolution, as shown by
\citet{mar83}. We checked this by formulating pure luminosity evolution, and
derived precisely the same results as for density evolution, the same minimum
value of the Likelihood function, and the same parameters, modified by the
relations to transcribe density into luminosity evolution given by \citet{mar83}.
Other forms of the redshift cut-off  were tried, in particular an exponential
roll-off:
\begin{equation}
\rho(L,z)=\rho_o(1+z)^k \exp[-(z/z_*)^n] l^{-\alpha}.
\end{equation}
The best fit for $n\,{=}\,1$ (Fig.~\ref{evol}, light blue lines) gives a minimum
likelihood markedly larger than that for the original form, while the best fits
for $n\,{=}\,2$ and $n\,{=}\,3$ are close in minimum likelihood value to the best
fit for the original form. Fig.~\ref{evol} shows why -- these latter two forms are
very similar, and are encompassed by the bootstrap results.  Note, however, that
these forms of roll-off introduce a fourth parameter.

To consider how the three missing redshifts in the total sample of 38 objects
might affect the reality of the cut-off, we adopted a conservative position: we
ascribed a redshift of 4.0 to each of the three unidentified sources in the total
sample of 38 objects. Running the minimization procedure for all 38 produced
(Table~\ref{tablum}; sample `complete+3')
the result shown as the dotted line in Fig.~\ref{evol}. It is encompassed by the
bootstrap trials; the missing redshifts do not change our conclusion. As a further
conservative test, we ran the minimization for the 24/35 objects with redshift
determinations from spectroscopy or other optical data. The resulting parameters
do not differ significantly from those for the sample of 35 objects; the cut-off is
secure.

There are two obvious ways in which our assumption of a single equivalent
temperature of 35K could be in error, a serious concern because of our choice of
rest-frame $850\,\mu$m as the luminosity measure. One is that there might be
significant scatter about this temperature. We tested the effect of this by a
simulation in which we adopted a Gaussian spread about 35K of $\sigma=10$K; our
particular simulation for the 35 objects yielded a maximum temperature of 59.6K
and a minimum of 15.4K. The resulting parameters are given in Table~\ref{tablum}.
They are encompassed within the errors for the complete sample. A second way in
which the equivalent temperature could differ from our single adopted value is a
dependence on redshift. We tried a dependence of the form $T(z) = 10 (1+z)\ K$
\citep{kov06}, and the parameters resulting (Table~\ref{tablum}) were
again unchanged within the uncertainties. It thus appears that we are viewing
predominantly {\it density or luminosity\/} evolution rather than spectral
evolution.

\begin{figure}
\vspace{12.0cm}

\includegraphics{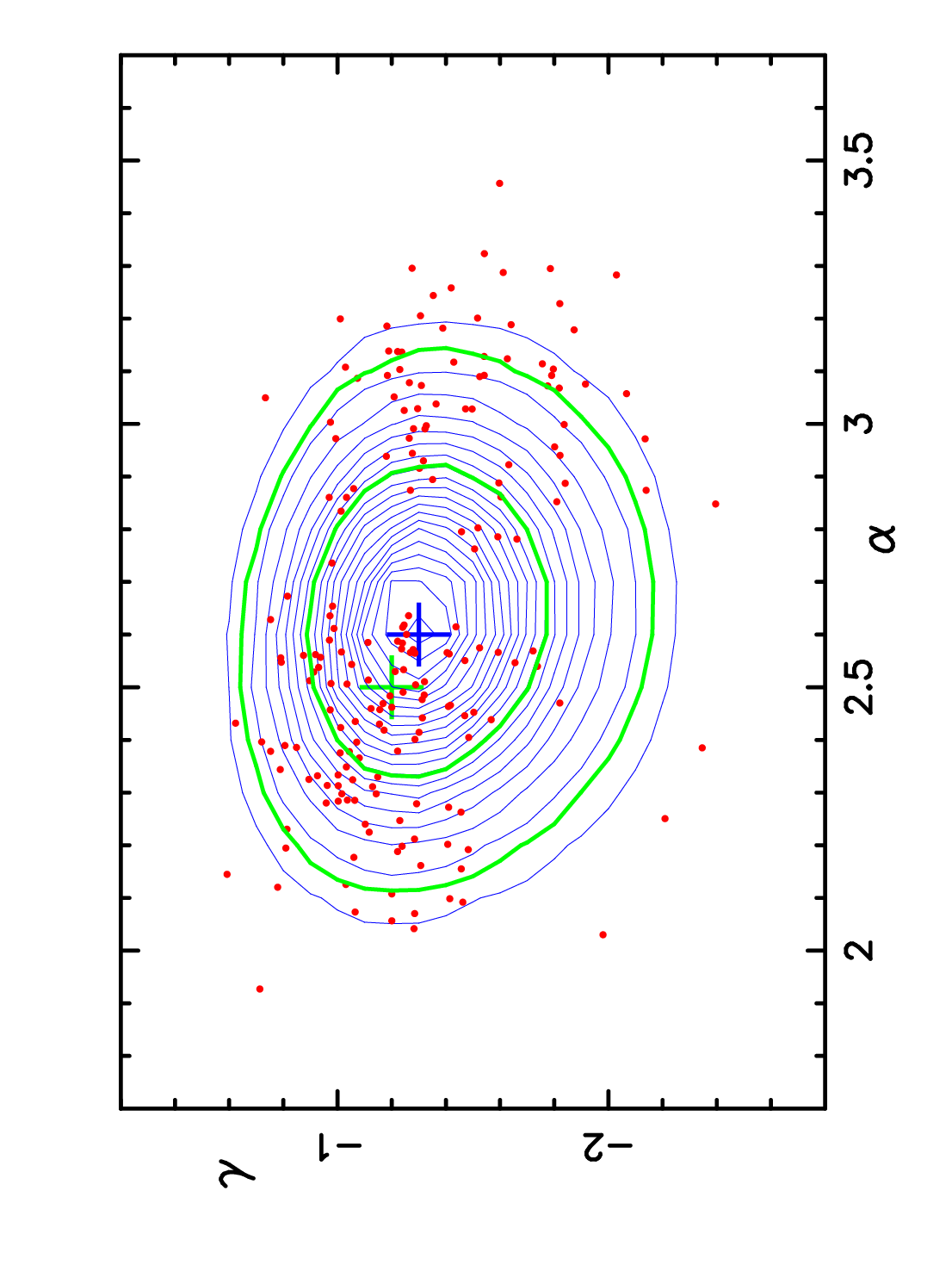}

\includegraphics{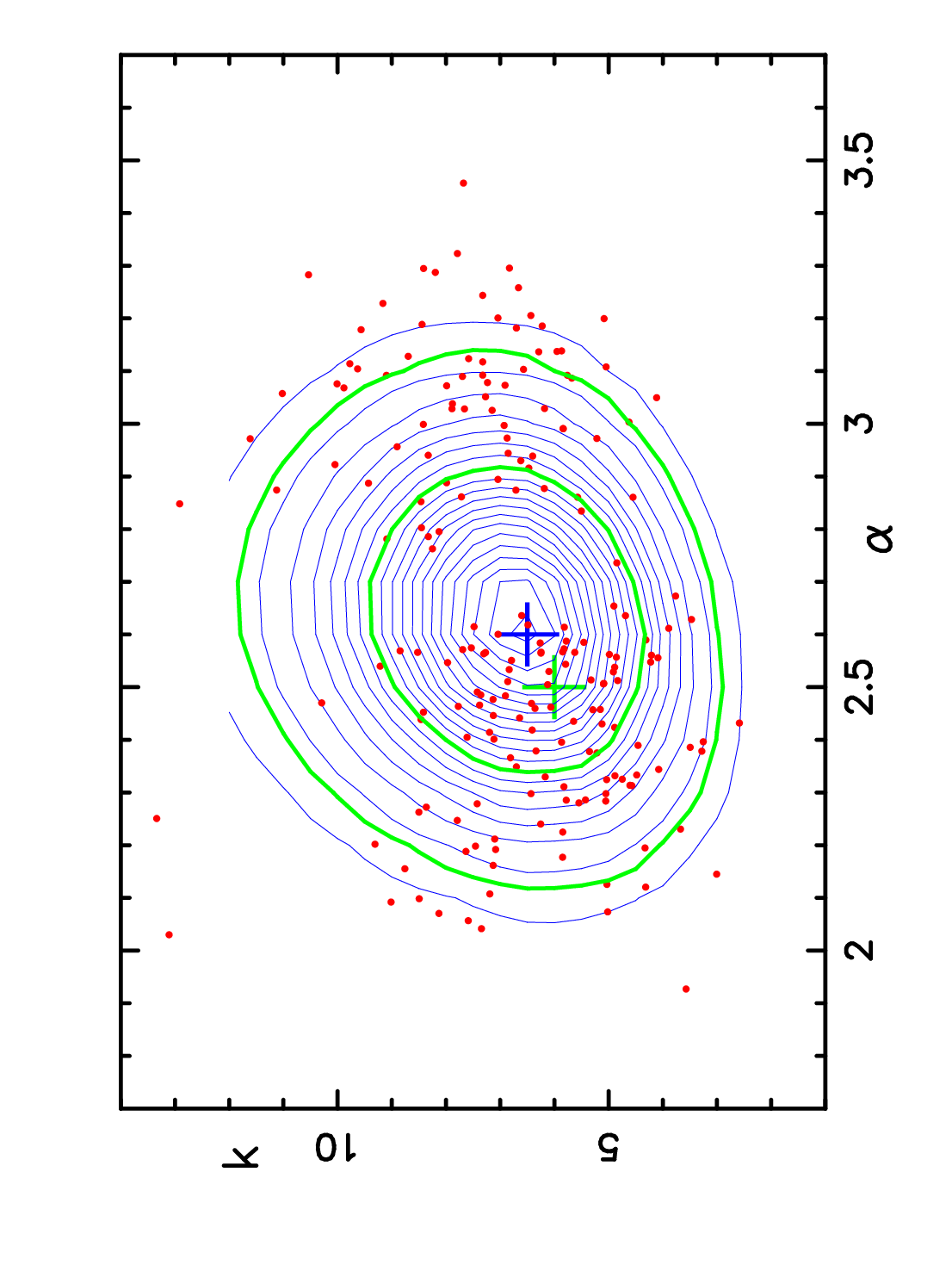}

\includegraphics{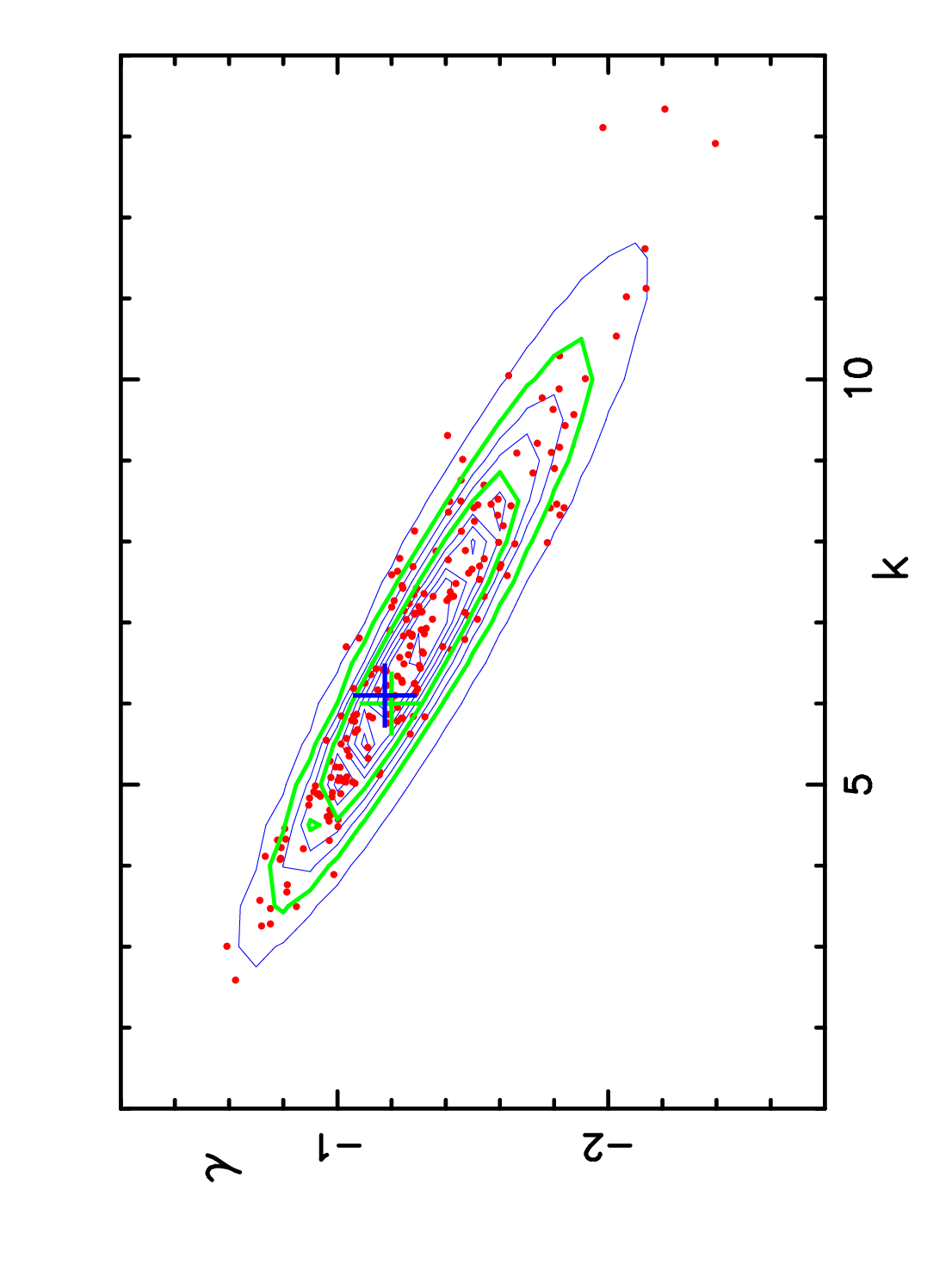}

\caption{Contours of probability for marginalized parameters of equation 8. The
contours are linearly spaced, with green contours representing 68 and 95 per cent
regions, respectively (corresponding to 1 and $2\sigma$, but for two-dimensional
data). Green crosses show the optimum values as determined by the downhill simplex
minimization, while blue crosses represent the marginalized contour minima. The
200 bootstrap results are plotted as dots. Note the `zones of avoidance' of the
bootstrap results, particularly in the central panel. This is discussed in
\S\ref{two}.}

\label{cons}

\end{figure}

\begin{figure}
\vspace{5.0cm}

\includegraphics{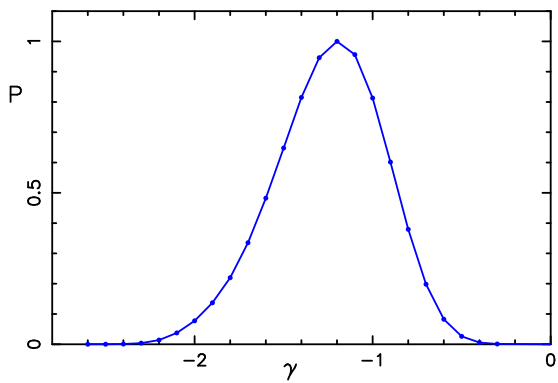}

\caption{The probability distribution for parameter
$\gamma$, which describes the deviation from redshift
power-law evolution through $(1+z)^{k+\gamma z}$. If
$\gamma$ is negative a redshift cut-off is implied, as the
$\gamma z$ term of the exponent must overpower $k$ at
redshifts somewhat greater than $z=-k/\gamma$.}

\label{pgam}

\end{figure}

The interplay between the parameters ($\alpha, k, \gamma$) can best be seen by
marginalization over each of them in turn, a process to examine degeneracies.
Fig.~\ref{cons} shows the marginalized posterior probability density functions for
pairs of the 3 parameters, assuming flat priors. The only degeneracy is the one
anticipated -- large values of $k$ (steep initial evolution) require
correspondingly large negative values of $\gamma$ to `restore' the space density
to its observed low values at high redshifts. There is no significant dependence
of the slope of the luminosity function on the evolution parameters. Marginalizing
over all parameters to find the probability distribution of $\gamma$ gives a clear
indication of the need for a redshift cut-off to describe the data. This
probability distribution is shown in Fig.~\ref{pgam}.
It indicates that there is essentially no probability of $\gamma$ being positive
-- the data demand a formulation of the evolving luminosity function which
specifies a redshift cut-off. For this model the significance of $\gamma \ge 0$ is
too small to measure.

\begin{figure}
\vspace{6.8cm}

\includegraphics{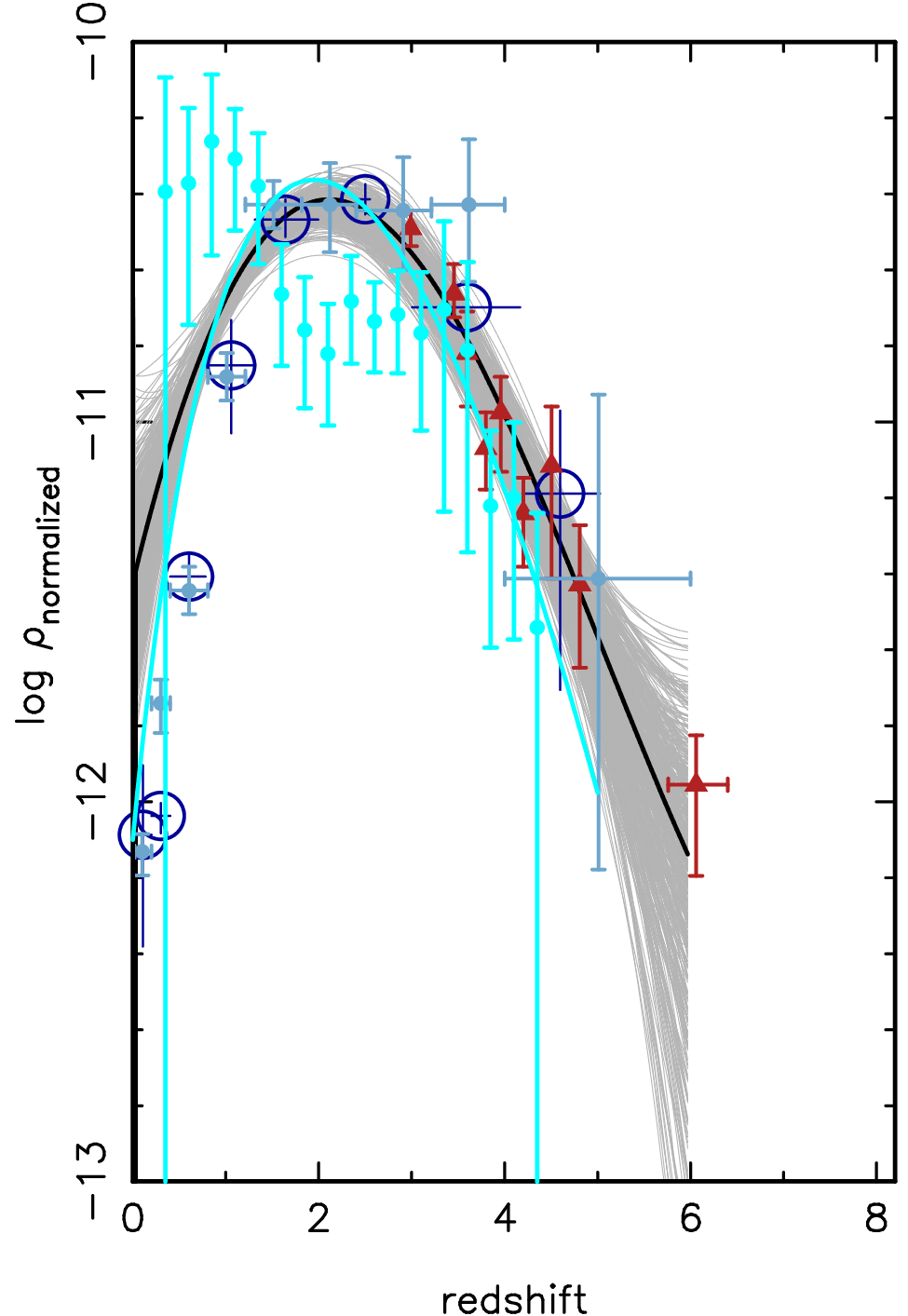}

\caption{The data of Fig.~\ref{evol} (light blue) superposed on a compilation (see
\citealt{wal05}) of QSO space-density dependences on redshift. The compilation
includes QSOs which are selected optically (red triangles: Sloan Digital Sky
Survey, \citealt{sch95,fan01b,fan04}), in X-ray bands (blue circles and crosses,
\citealt{has04,has05,sil05}) and from radio surveys (grey shading, black line,
\citealt{wal05}).}

\label{sm-qso}

\end{figure}

Fig.~\ref{sm-qso} shows a comparison of the epoch behaviour of the luminosity
function with the space-density dependence established for QSOs selected in
different wavebands. The coincidence in form is remarkable, and is discussed
further in \S\ref{discon}.

The bootstrap results in this analysis indicate broad
agreement with the simple adopted model, but do not
inspire confidence in either the model details or the
parameters derived from it. The bootstrap parameter
distributions are non-Gaussian and show zones of
avoidance, particularly in the upper and middle panels of
Fig.~\ref{cons}. We return to this issue in \S\ref{two}.

\section{The star formation rate}

\label{sfr}

From the spectral assumptions of a greybody with $\beta=1.5$ and $T=35\,$K, we
calculated the total IR luminosity and converted it to star formation rate for
each galaxy using the relationship for starburst galaxies given by \citet{ken98}.
This `cold-dust' SFR assumes a \citet{sal55} initial mass function and applies to
starbursts with ages less than $100\,$Myr. It also assumes little or no AGN
contribution to the IR luminosity, a reasonably good assumption for SMGs
\citep{ale05a,pop06}. Furthermore, since we have assumed only a greybody template
with one temperature, we are not including any contribution of warm dust and/or
mid-IR spectral features to the IR luminosity -- the values we use here are for
the cold dust only, which is expected to dominate in such systems
{\citep{pop06,huy06}. Because of these and other systematic effects, our results
will be difficult to compare in detail with SFRD derived from samples selected in
other wavebands. Nevertheless, the results should give reasonable estimates for
SFRD evolution, provided that the dust properties do not vary appreciably with
redshift.

Dividing space up into redshift shells, the volume
contribution for each galaxy was calculated from $\Delta V
= V_{\rm max} - V_{\rm min}$, where $V_{\rm min}$ is the
lower redshift limit of the shell, and $V_{\rm max}$ is
either the shell upper redshift limit or the $V_{\rm max}$
value determined from the redshift at which the galaxy
encounters its individual survey limit line
(Fig.~\ref{Pz-submm1}) -- whichever is smaller. Each
galaxy then makes a contribution to the SFRD in the shell
of $({\rm SFR}_i/\Delta V_i)\times(4\pi/A_i)$ where $A_i$
is the area of each `single-source-survey', as described
earlier.  The result of such a calculation for redshift
shells of $\Delta z = 0.6$ is shown in Fig.~\ref{sfr-z} --
of course the results are not very different from a scaled
version of Fig.~\ref{evol}.

\begin{figure}
\vspace{5.0cm}

\includegraphics{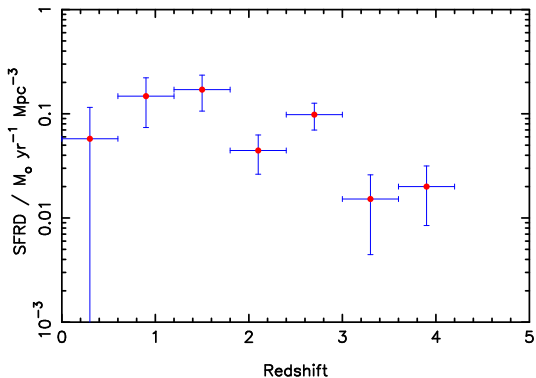}

\caption{Star formation rate density as a function of redshift,
in redshift shells $\Delta z = 0.6$.}

\label{sfr-z}

\end{figure}

Although these estimates are noisy, there is evidence from the plot that the SFRD
from SMGs declines at redshifts beyond three. Fig.~\ref{sm-sfr} shows the points
of Fig.~\ref{sfr-z} in comparison with numerous other recent estimates of the
epoch dependence of star formation rate density.

\begin{figure}
\vspace{7.0cm}

\includegraphics{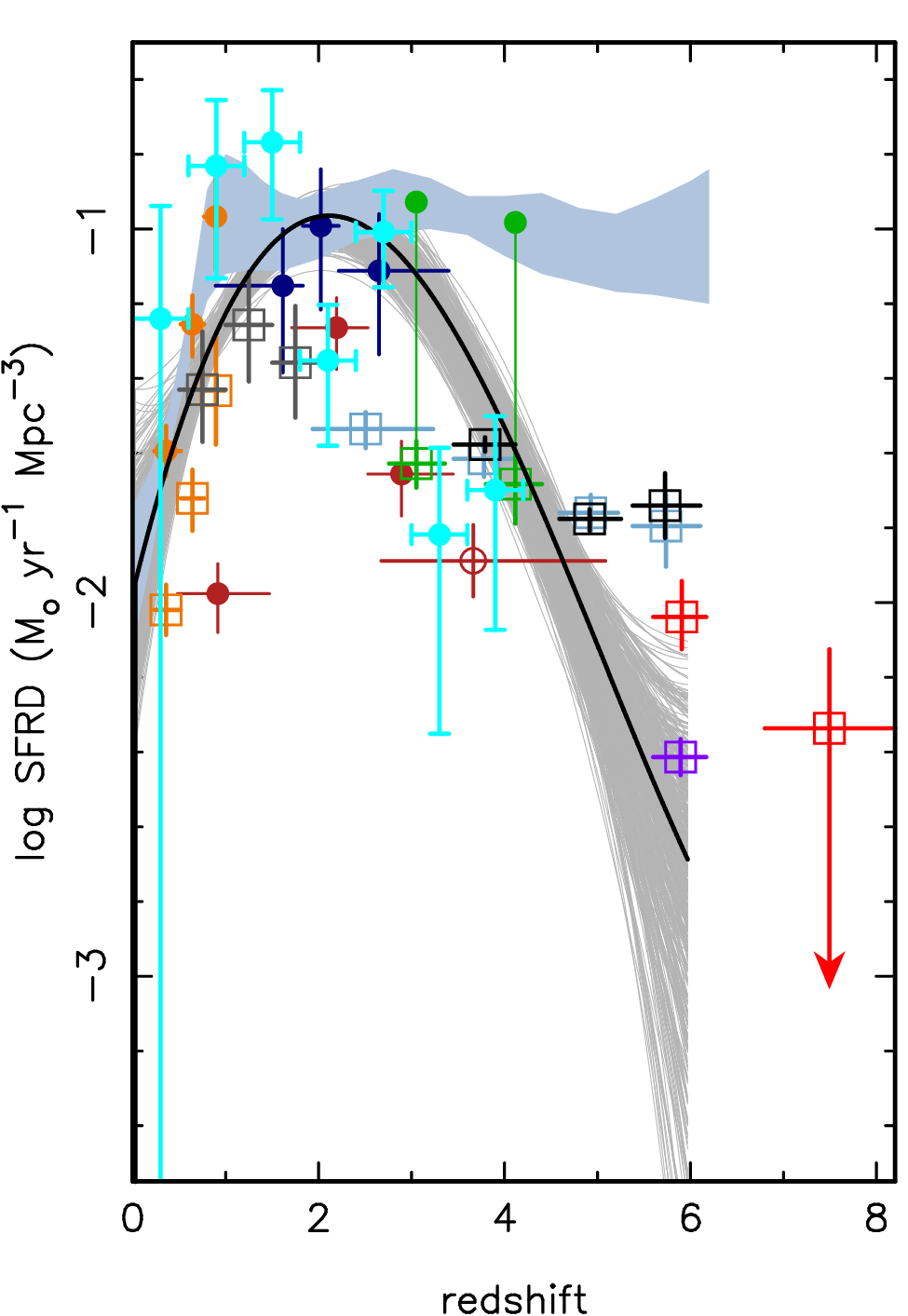}

\caption{The data of Fig.~\ref{sfr-z} (light blue)
superposed on a compilation of SFRD estimates as a
function of redshift; see \citet{wal05} for details. The
black line and grey shading represents the space
distribution of radio QSOs, from Fig.~\ref{sm-qso},
arbitrarily scaled. The dark red circles \citep{cha05}
represent perhaps the most important comparison; these are
estimates from a sample of radio-detected,
spectroscopically-confirmed SMGs, but are not an unbiased
sample as represented by the present data. }

\label{sm-sfr}

\end{figure}

\section{Two populations?}
\label{two}

The unsatisfactory bootstrap results shown in Fig.~\ref{cons} suggest
non-Gaussianity and more specifically a dichotomy, indicating that we may be
looking at two populations.

We originally suspected this feature of the population to be due to sample
problems, from errors in redshift estimates, errors in identifications, or to a
unique distribution of luminosities whose analysis would be fragile in the face of
errors in the identification process. We therefore tested for robustness in a
number of different ways.

We first considered errors in the redshift estimates. Assuming the spectroscopic
redshifts to be correct, we ran many realizations of our sample with the
photometric redshifts randomly dispersed about the measured values, applying
Gaussian errors of $\Delta z = 0.14(1+z)$, twice the size of the error estimated
in these redshifts by \citet{pop06}. The bootstrap tests always yielded plots of
similar appearance, with a bifurcated spread of points. We then dropped the
assumption that the spectroscopic redshifts were correct, and repeated the
exercise for all objects in the sample. The parameter set for the first of these
realizations is in Table~\ref{tablum}, its values encompassed again by the spread
in values from the original sample. The results persisted, the same bifurcated
spread of points appearing. Thus the bootstrap structure does not appear to be the
result of `preferred' redshift estimates.

We modelled errors in identifications in two ways. First we reduced the sample by
randomly throwing out `misidentifications', creating sample realizations by
removing one, two or three objects at random in the following sequence. For one
object, a single SMG was removed from the set of 12 less secure identifications;
for two objects, one was removed from the 23 secure identifications and one from
the 12 less secure; for three objects, two were removed from the less secure
identifications and one from the secure identifications. Three errors in
identifications is the maximum number we would anticipate, based on the summed
probabilities of identification reliabilities.

Secondly we retained the sample size and mimicked the possibility of
misidentifications as follows. We assumed that the redshift distribution is
correct; this is a reasonable approximation, as the great majority of sample
members are correctly identified and there is no identifiable bias in the redshift
determinations. We then assumed one, two or three objects at random were
misidentified, following the foregoing choice sequence amongst the secure and less
secure identifications. We assigned a new redshift to each of these one, two or
three objects, drawn at random from the distribution of all redshifts in the
sample.

A maximum of three identification errors might be realistically expected from the
summed identification probabilities. To be extreme (and unrealistic), we tried
simulations in which up to 8 of the identifications were deemed incorrect.

None of these realizations destroyed the general appearance of the bootstrap
results shown in Fig.~\ref{cons}. In some cases the zone of avoidance was less
well defined; but all showed the same relatively tight clustering of points to the
left of the contour maximum, and a rather more diffuse distribution of points to
the right (upper two panels, Fig.~\ref{cons}). In no case did the general
appearance of `two clumps' disappear.

Although this was equally true for the realizations involving excess redshift
errors, in these cases the errors had the effect of reducing the evolution
parameter $k$ (Table~\ref{tablum}), shifting the distribution downward in the
central panel of Fig.~\ref{cons}. This is understandable in that adding noise to
the redshift distribution (see Fig.~\ref{Pz-submm1}) will reduce the sharpness of
the `rise' towards $z = 1$ and make the fall-off after $z=4$ somewhat more gentle.
The effect is equivalent to broadening the distribution of luminosities in the
complete sample. Less space-density evolution will be required. Placing a scatter
on the equivalent temperature produces the same broadening of the luminosity
distribution, and again the lower value of the evolution exponent is evident in
Table~\ref{tablum}.

A bifurcated spread of points as a stable feature of the sample suggested that the
data have something more to tell us.

To examine this, we divided the sample of 35 at the median luminosity of
log~$L_{850 \mu m} = 23.2$ and repeated the likelihood analysis for each
subsample. The minimization routine yielded the results set out once more in
Table~\ref{tablum}.

The differences between the parameters for the subsamples now strongly suggest the
presence of two distinct populations. Although the individual parameters do not
differ at high significance levels, the joint probabilities show the subsamples
occupying distinct and markedly different regions of the 3D parameter space. The
lower-luminosity objects show a luminosity function of slope around $-2$ and
relatively mild cosmic evolution, $k \sim 5$. The slope of the luminosity function
for the more luminous objects is closer to $-3$ and the evolution is much more
dramatic, with $k \sim 14$. In both subsamples, the value of $\gamma$, the
redshift cut-off parameter, is significantly below zero; the data support a
redshift cut-off for each subsample.

A global minimization solution of the likelihood function for two sub-populations
yielded essentially identical results. In this test each sub-population was
required to have independent evolution described by the three parameters, with the
dividing luminosity set as a seventh free parameter. The key point from this
analysis is that the luminosity split between the sub-populations emerged as
identical (to 1 decimal place in the log) to the log median adopted {\it a
priori}.

The previous probability analyses with marginalizations were then carried through
for the two subsamples individually.  The structure of the points in
Fig.~\ref{evol} (reproduced in Fig.~\ref{evol2}) is better represented by the
two-component luminosity function, as Fig.~\ref{evol2} demonstrates.

\begin{figure}
\vspace{5.0cm}

\includegraphics{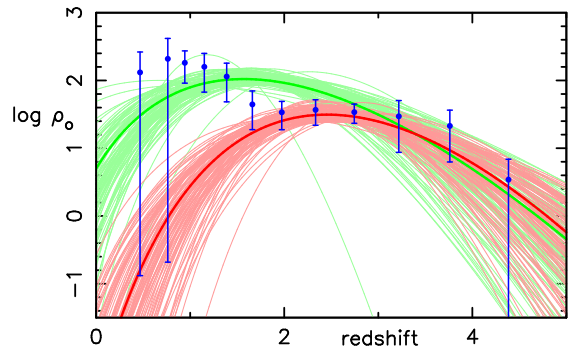}

\caption{The points and error bars are the data described and presented in
Fig.~\ref{evol}. The green curve is a functional fit to the low-luminosity data
using maximum likelihood with a minimization routine, and faint green lines
represent 100 bootstrap trials of the evolution model, with bootstrapping
end-to-end from this subsample of 17 SMGs. The red curve and its faint red
counterparts are the best fit and bootstrap fits for the 18 higher-luminosity
objects. The sum of the two components is a reasonable representation of the form
of the successive redshift-slice values of $\rho_0$.}

\label{evol2}

\end{figure}

The three diagrams of Fig.~\ref{cons} now become 6 diagrams; as representatives,
Fig.~\ref{cons2} gives the two separate $k-\alpha$ plots, the plane (middle panel,
Fig.~\ref{cons}) which previously showed the least satisfactory distribution of
bootstrap points. The bootstrap points in these two plots are now distributed as
expected, suggesting Gaussianity prevails for each sub-sample.

\begin{figure}
\vspace{8.0cm}

\includegraphics{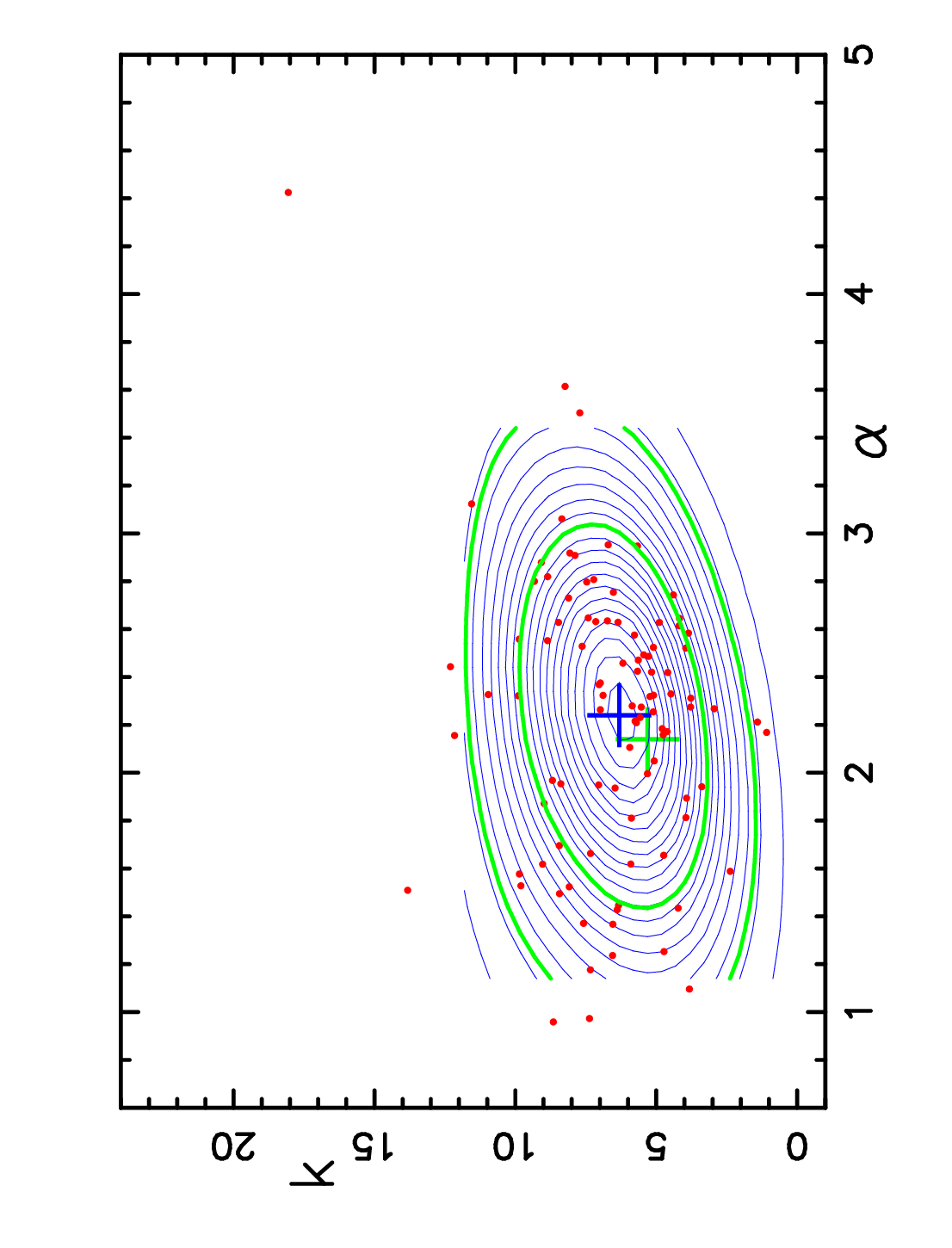}

\includegraphics{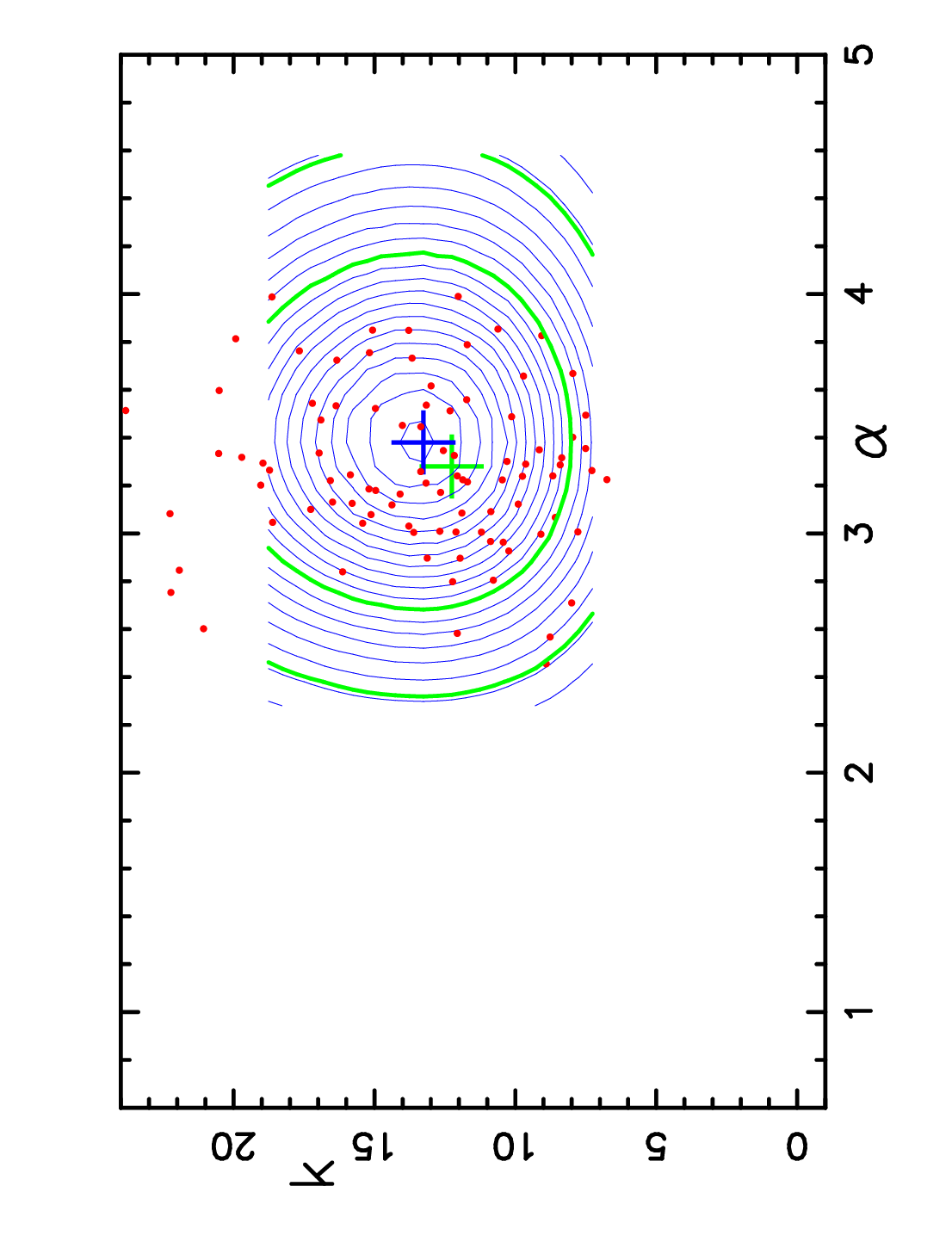}

\caption{Contours of probability for marginalized
parameters $k$ (evolution as $(1+z)^k$) and $\alpha$
($-$slope of the power-law luminosity function): top,
low-luminosity subsample; bottom, high-luminosity
subsample. The plots are on the same axes to demonstrate
that the preferred regions hardly intersect, i.e.~both parameters differ
between the two subsamples. As before the contours are
linearly spaced: green contours represent 68 per cent and
95 per cent regions, respectively. The distributions of
the bootstrap points are now approximately Gaussian in
appearance; in each case $68 \pm 2$ out of the 100 points
fall within the effectively $1\sigma$ contours. }

\label{cons2}

\end{figure}

To demonstrate the significance of the difference between
the two sub-populations, we computed the value of the
likelihood function for each subsample using the
maximum-likelihood parametric fit obtained for the other
subsample. We then compared this value with the
maximum-likelihood value found for the subsample. The
respective differences for the low-luminosity sample and
the high-luminosity sample were 25.6 and 19.0 in $\chi^2$
for 3 degrees of freedom. This indicates rejection of the
model for each subsample by the data of the other
subsample, at the $\le 0.001$ level of significance. In
addition we ran 1000 bootstrap tests on each subsample to
find how frequently the resultant model parameters
overlapped those determined from the maximum-likelihood
solution for other subsample. Considering for example the
\textsl{high-luminosity sample} bootstraps, how many times
out of 1000 would we find (see Table~\ref{tablum}) $\alpha
\le 2.1, k \le 5.3$ {\it and} $\gamma \ge -1.3$? In fact
we found $0/1000$, and for the \textsl{low-luminosity
sample} we found $3/1000$. This test again indicates a
difference between the subsample models at a significance
level of about $0.001$. (Note that these tests are valid
{\it only} because we split our sample into high and
low-luminosity subsamples {\it a priori}, i.e.~without
optimization.)

The single-dimensional distribution of SMG luminosities
shows no strong indication of a dichotomy.
However, this is not an argument against the presence of two
populations; samples of 100s of radio sources likewise
show no clear dichotomy in the luminosity distribution,
despite the known presence of low-luminosity and
high-luminosity populations, largely distinct in
morphology and evolving very differently
\citep{dun90,jac99,sad06}. Beyond the inevitable
correlations of flux and luminosity with redshift,
\citet{pop06} found no additional correlations of spectral
properties with redshift.

We conclude that statistically the two-population hypothesis is
solidly based. However we cannot deny that the sample is small, with the remote
possibility of a unique assemblage of objects leading us astray.

Finally, we carried out the SFR calculation individually for the two subsamples.
The procedure as described in \S\ref{sfr} was followed, with the results appearing
in Fig.~\ref{sfr-z2}.
\begin{figure}
\vspace{5.5cm}

\includegraphics{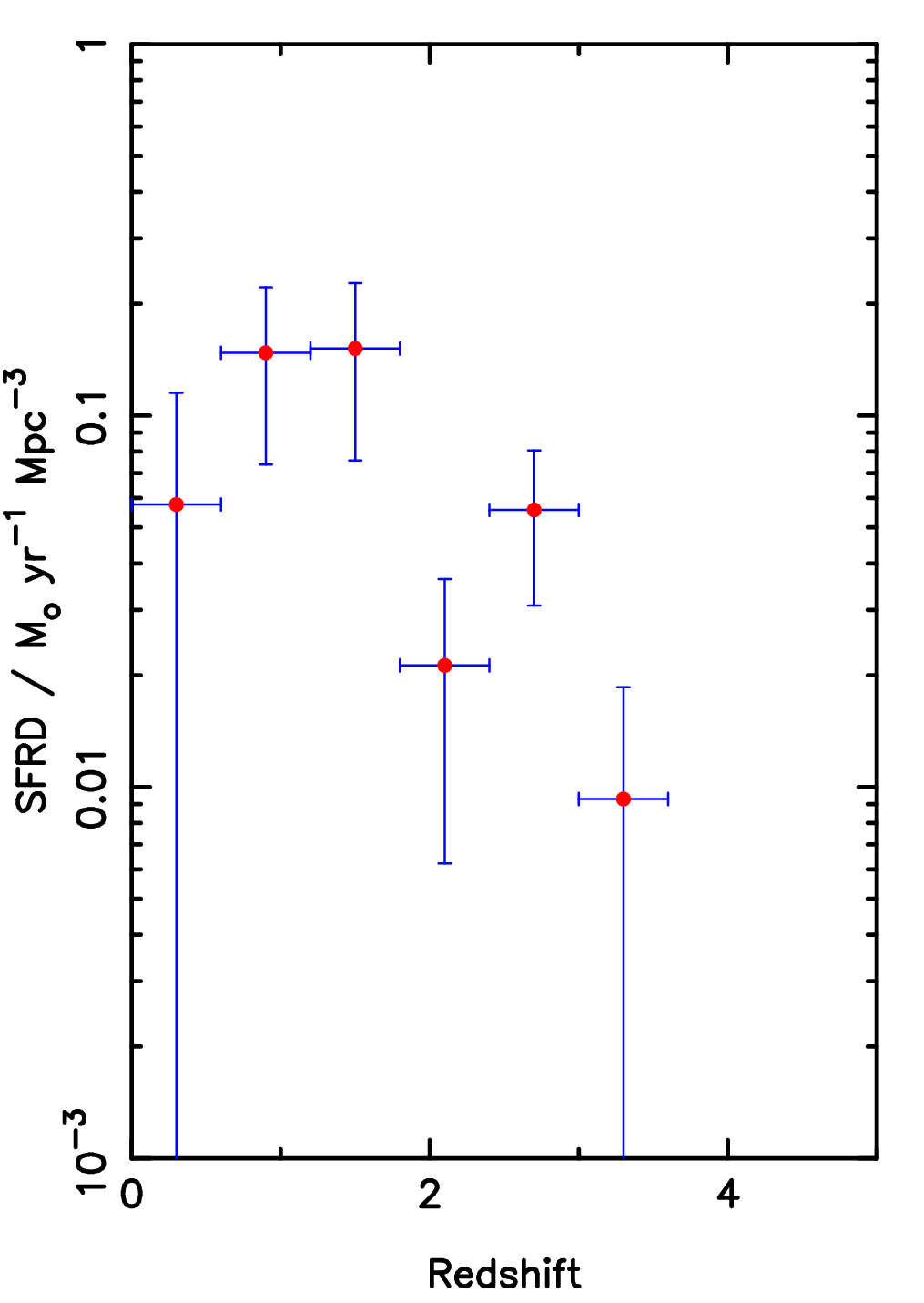}

\includegraphics{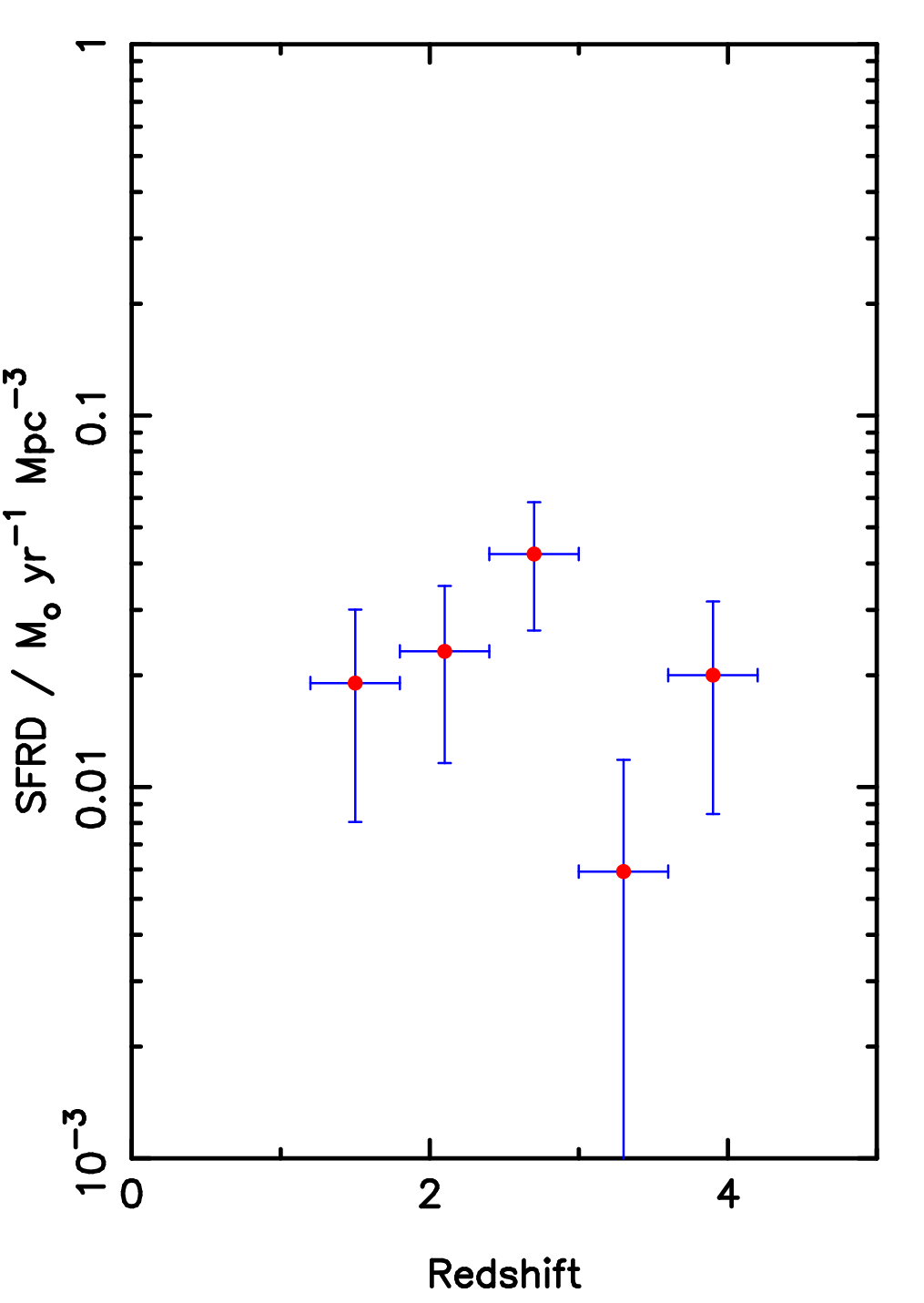}

\caption{Star formation rate density as a function of redshift, using redshift
shells $\Delta z = 0.6$: left-hand panel, low-luminosity SMG subsample; right-hand
panel, high-luminosity SMG subsample.}

\label{sfr-z2}

\end{figure}
The diagram shows that the star-formation rate dependence
on epoch differs for the two sub-populations, the SFRD
peaking around $z\sim 1.5$ for the lower luminosities and
around $z\sim 2.5$ for the higher luminosities. Of course
this result is not at all independent of the different
forms of SMG volume-density evolution found for the two
subsamples (Table~\ref{tablum} and Fig.~\ref{evol2}).

\section{Discussion and Conclusions}
\label{discon}

We have shown that there is a significant decline in the space density of SMGs
beyond a redshift of three.  This conclusion has undergone extensive testing via
bootstrap analyses plus the investigation of different forms for the evolution.

Several authors \citep[e.g.][]{san88,gen98,arc02,ste05,dim05,ale05b} have
suggested a connection between the formation of powerful QSOs and ULIRGs (or their
high-$z$ counterparts the SMGs). A popular picture has emerged of an evolutionary
sequence in which the forming galaxy is initially far-IR luminous but X-ray weak,
similar to the sources discovered as SMGs.  As the black hole and spheroid grow
with time, a point is reached when the central QSO becomes powerful enough to
terminate the star formation and eject the bulk of the fuel supply.  This
transition is followed by a period of unobscured QSO activity, subsequently
declining to leave a quiescent spheroidal galaxy. For the first time
(Fig.~\ref{sm-qso}) we have been able to compare the space densities of QSOs and
SMGs. Such a scenario is consistent with our results, in which we find remarkable
concordance between the space density decline shown by the SMGs, by all types of
QSOs and by the SFRD from SMGs. Examining the significance of the time sequence is
beyond the capabilities of the present data, but at a minimum the data emphasize
the strong connection between SMGs, AGN activity and cold-dust SFRD.

The cold-dust-derived SFRD from SMGs shows a significant decline at redshifts
beyond about three.  The larger star-formation rate from the more distant and
higher-luminosity objects is inadequate to overcome the rapid decline in their
volume density.  If there is significant star formation beyond redshifts of 4, it
is not the province of SMGs, but must be carried by different and generally
lower-luminosity populations, such as the Lyman-break galaxies \citep{ste99} or
galaxies found in very deep searches at optical wavelengths \citep{bou04,gia04}.
Semi-analytic modelling of the SFRD from SMGs \citep[e.g.][]{bau05} suggests a
broad peak at $2<z<3$, although the predicted diminution to higher redshifts is
less than that indicated by the results here.

There is evidence suggesting that the submm population may be strongly clustered
\citep[e.g.][]{bla04}. Although our results are based on submm observations in a
small field, we note that our main conclusions will be unaffected by clustering
unless the clustering strength of SMGs depends strongly on their luminosity.
However, the redshifts of the sample members are so large and so diverse that
there is no possibility of sample members occupying the same supercluster or
filament. There is thus little likelihood that cosmic variance is affecting the
results.

The data are remarkably insistent on the presence of two subpopulations of
objects, divided by luminosity. These evolve in distinctly different ways and
their luminosity functions have different shapes. Their SFRD histories are
likewise very different. The ULIRG/LIRG dichotomy is of particular relevance here,
and our results are similar to those discussed in some earlier studies of lower
redshift populations \citep[e.g.][]{kim98,gui98,cha01,lag03,saj03,xu03}, sometimes
more loosely described as a distinction between `starbursts' versus more normal
galaxies. At higher redshifts \citet{cha06} illustrated (in his fig.~4) how SFRD
dominance shifts from ULIRGs at $z\ge2.5$ to LIRGs at $z\sim 1$ (see also
\citealt{cap07} and other {\sl Spitzer}-based studies).  Our dividing line in
luminosity is somewhat more extreme than the LIRG/ULIRG boundary, normally taken
at $10^{12} {\rm L}_\odot$; our division at $\log L_{850 \mu\rm{m}} = 23.2$
corresponds to about $3\times10^{12}{\rm L}_\odot$.
Despite our higher adopted dividing line, our results parallel those for
LIRGs/ULIRGS: we find the most IR-luminous SMGs dominating the energy output (or
SFRD) at $z\sim2.5$, while the less luminous SMGs dominate the SFRD at $z\sim1$.
Note that this analysis does account for the contribution to the SFRD from high
redshift LIRGs and ULIRGs selected at shorter IR or radio wavelengths. Within the
submm population, we are seeing a down-sizing in the luminosity of the dominant
contributors to the energy budget.


We conclude that a redshift cut-off is established for SMGs in both object density
and SFRD, both of which are similar in form to cut-offs found for powerful AGN.
The redshift cut-off is established with such certainty that that for the models
adopted, the level of significance is too small to calculate. We also conclude
that at a level of significance of $\sim0.001$, two populations are probably
present amongst SCUBA-detected SMGs, showing distinctly different evolutionary
histories and luminosity functions. Although cosmic dowsizing is certainly
present, we note that our sample is small; and despite extensive testing, the
two-population hypothesis could have resulted from a singular grouping of data.
However,  we can be optimistic that with much larger samples soon to be collected
using SCUBA-2 \citep{hol06}, it will be possible to test the several ideas and
issues which arise from our study.  These include: further probing of the two
population question; the AGN--SFR connection and its time-lag; the role merging
plays in this process; details of how and why SMGs possibly organize themselves to
manifest cosmic down-sizing; and the relation this down-sizing has to populations
of LIRGs and ULIRGs selected at other wavelengths.

 \vspace{4mm} \noindent{\it Acknowledgements.} We are very grateful
to Chris Blake for many helpful discussions and suggestions, and to Mitch Crowe
for work on including additional `two bolometer chopping' SCUBA data in the
supermap.  Helpful comments from a referee resulted in a significantly improved
manuscript. This work was supported by the Natural Sciences and Engineering
Research Council of Canada and by the Canadian Space Agency.

\appendix
\section{Data for the three additional sources of the sample}
\samepage\begin{table*}
\begin{center}
\caption{The three new GOODS-N supermap sources.}
\begin{tabular}{ccccccll}
\hline\hline Submm ID & Submm name & Radio RA & Radio Dec& Raw S$_{850}$
(mJy)&Deboosted S$_{850}$ (mJy)
& Redshift & P\\
\hline
& &&&                  & &   &   \\
GN39&SMMJ123711.1+621325&12:37:11.33&62:13:31.02&7.4$\pm$1.9&5.2$\pm$2.4&\ \ 1.996&0.02\\
&&12:37:11.97&62:13:25.77&&&&\\
     &&&        &           &          &     &          \\
GN40&SMMJ123713.7+621822&12:37:13.86&62:18:26.24&13.1$\pm$2.7&10.7$\pm$2.9&\ \ 2.6&0.003\\
     &&&        &           &          &     &          \\
GN41&SMMJ123639.4+620752&n/a&n/a&11.9$\pm$3.1&8.8$\pm$3.5&&\\
\hline
\end{tabular}
\label{tab_newdata}
\end{center}
\end{table*}
Table~\ref{tab_newdata} gives positions, fluxes and redshifts of the three new
GOODS-N supermap sources. The submm name gives the position of the submm source
while RA and Dec provide the position of the radio counterpart.  The counterpart
identifications were performed exactly as outlined by \citet{pop06}. We list both
the raw and deboosted submm fluxes. $P$ is the probability that the counterpart is
a random association [see \citet{pop06} for more details on these probabilities].
GN39 appears in the \citet{cha05} catalogue and has two radio counterparts, both
confirmed to lie at the same redshift \citep{swi04,cha05}; we list the two radio
positions. The redshift of GN40 is an IRAC-only photometric redshift as described
by \citet{pop06}. GN41 does not have a unique likely counterpart and is therefore
not included in the analysis of this paper.

\end{document}